\documentclass[aps,pra,amsmath,amssymb,reprint,superscriptaddress,floatfix]{revtex4-1}

\usepackage{xcolor}
\usepackage[]{graphicx}
\usepackage{xspace}
\usepackage{dsfont}
\usepackage{hyperref}

\newcommand{\red}[1]{#1} 

\newcommand{\ket}[1]{\vert #1 \rangle} 
\newcommand{\ketbra}[2]{\left| #1 \middle> \! \middle< #2
\right|} 
\newcommand{\abs}[1]{\vert #1 \vert} 
\newcommand{\avg}[1]{\langle #1 \rangle} 
\newcommand{\I}{\mathrm{i}} 
\newcommand{\unit}[1]{\ensuremath{\, \mathrm{#1}}}
\newcommand{\commutator}[2]{[ #1 , #2 ]} 
\newcommand{\br}[1]{\left( #1 \right)}
\newcommand{\brr}[1]{\left[ #1 \right]}
\newcommand{\brrr}[1]{\left\{ #1 \right\}}

\newcommand{\de}{\mathrm{d}}
\newcommand{\td}[2]{\frac{\de #1}{\de #2}} 

\DeclareMathOperator{\Tr}{Tr}
\DeclareMathOperator{\sech}{sech}

\newcommand{\Lio}{\mathcal{L}} 

\newcommand{\Ein}{\mathcal{E}_\mathrm{in}}

\newcommand{\Eout}{\mathcal{E}_\mathrm{out}}
\newcommand{\E}{\mathcal{E}}
\newcommand{\F}{\mathcal{F}}
\newcommand{\Tc}{T_\mathrm{c}}
\newcommand{\MHz}{\times2\pi\unit{MHz}}
\newcommand{\wc}{\omega_\mathrm{c}}
\newcommand{\wl}{\omega_\mathrm{L}}
\newcommand{\kbad}{{\kappa_\mathrm{loss}}}

\newcommand{\OF}{\Omega^\mathrm{F}}
\newcommand{\OG}{\Omega^\mathrm{G}}
\newcommand{\OD}{\Omega^\mathrm{D}}
\newcommand{\OX}{{\Omega^\mathrm{X}}}
\newcommand{\Oo}{\Omega^\mathrm{opt}}

\begin{document}

\title{Optimal storage of a single photon by a single intra-cavity atom}

\author{Luigi Giannelli} \affiliation{Theoretische Physik, Universit\"at des
  Saarlandes, 66123 Saarbr\"ucken, Germany}

\author{Tom Schmit} \affiliation{Theoretische Physik, Universit\"at des
  Saarlandes, 66123 Saarbr\"ucken, Germany}

\author{Tommaso Calarco} \affiliation{Institute for Complex Quantum Systems \&
  Centre for Integrated Quantum Science and Technology, Universit\"at Ulm,
  89069 Ulm, Germany}

\author{Christiane P. Koch} \affiliation{Theoretische Physik, Universit\"at
  Kassel, Heinrich-Plett-Str. 40, 34132 Kassel, Germany}

\author{Stephan Ritter} \altaffiliation[present address: ]{TOPTICA Photonics
  AG, Lochhamer Schlag 19, 82166 Graefelfing, Germany}

\affiliation{Max-Planck-Institut f\"ur Quantenoptik, Hans-Kopfermann-Strasse 1,
  85748 Garching, Germany}

\author{Giovanna Morigi} \affiliation{Theoretische Physik, Universit\"at des
  Saarlandes, 66123 Saarbr\"ucken, Germany}


\date{\today}

\begin{abstract}
  We theoretically analyse the efficiency of a quantum memory for single
  photons. The photons propagate along a transmission line and impinge on one
  of the mirrors of a high-finesse cavity. The quantum memory is constituted by
  a single atom within the optical resonator. Photon storage is realised by the
  controlled transfer of the photonic excitation into a metastable state of the
  atom and occurs via a Raman transition with a suitably tailored laser pulse,
  which drives the atom. Our study is supported by numerical simulations, in
  which we include the modes of the transmission line and we use the
  experimental parameters of existing experimental setups. It reproduces the
  results derived using input-output theory in the corresponding regimes and can
  be extended to compute dynamics where the input-output formalism cannot be
  straightforwardly applied. Our analysis determines the maximal storage
  \red{efficiency}, namely, the maximal probability to store the photon in a
  stable atomic excitation, in the presence of spontaneous decay and cavity
  parasitic losses. It further delivers the form of the laser pulse that
  achieves the maximal \red{efficiency} by partially compensating parasitic
  losses. We numerically assess the conditions under which storage based on
  adiabatic dynamics is preferable to non-adiabatic pulses. Moreover, we
  systematically determine the shortest photon pulse that can be efficiently
  stored as a function of the system parameters.
\end{abstract}


\maketitle

\section{\label{sec:introduction}Introduction}
Quantum control of atom-photon interactions is a prerequisite for the
realization of quantum networks based on single photons as flying qubits
\cite{Cirac1997,Kimble2008}. In these architectures, the quantum information
carried by the photons is stored in a controlled way in a stable quantum
mechanical excitation of a system, the quantum
memory~\cite{Afzelius2015,Englund2016,Kalachev2007,Kalachev2008,Kalachev2010}.
In several experimental realizations the quantum memory is an ensemble of spins
and the photon is stored in a spin wave excitation~\cite{Afzelius2015}.
Alternative approaches employ individually addressable particles, such as
single trapped atoms or ions~\cite{Duan2010,Reiserer2015}: here, high-aperture
lenses~\cite{Kurz2014} or optical resonators~\cite{Ritter2012} increase the
probability that the photon qubit is coherently transferred into an electronic
excitation. \red{In addition, schemes based on heralded state transfer have
  been realized~\cite{Kurz2014,Kalb2015,Kurz2016,Brito2016}}. Most recently,
storage \red{efficiencies} of the order of 22\% have been reported for a
quantum memory composed by a single atom in an optical
cavity~\cite{Koerber2017}. This value lies well below the value one can extract
from theoretical works on spin ensembles for photon
storage~\cite{Gorshkov2007}. This calls for a detailed understanding of these
dynamics and for elaborating strategies to achieve full control of the
atom-photon interface at the single atom level.
\begin{figure}[!ht]
  \centering \includegraphics[width=\linewidth]{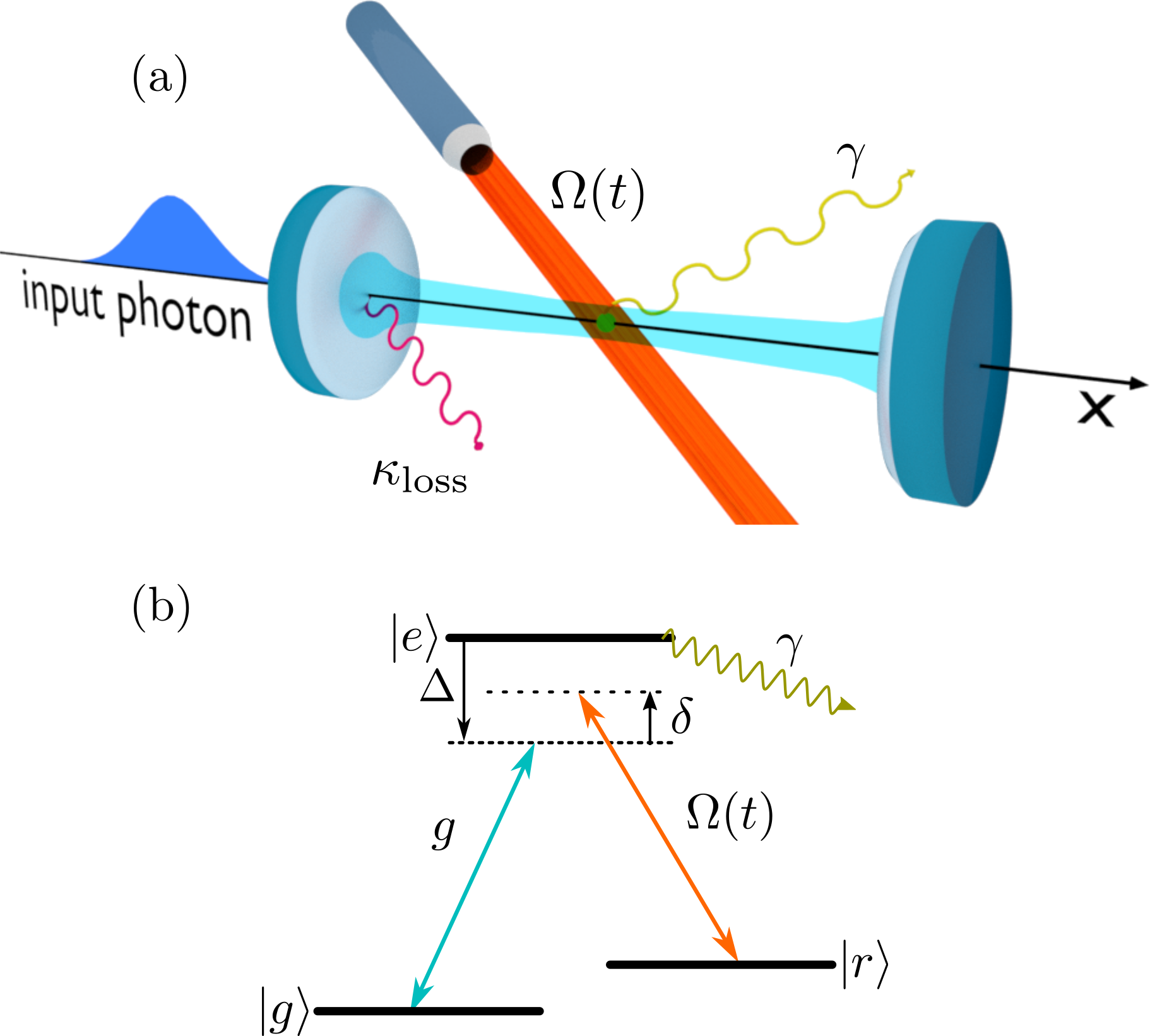}
  \caption{\label{Fig:1}Storage of a single photon in the electronic state of a
    single atom confined inside an optical resonator. (a) The photon wave
    packet propagates along a transmission line and impinges onto a cavity
    mirror. (b) The single photon is absorbed by the cavity, which drives the
    atomic transition $|g\rangle \to |e\rangle$. An additional laser couples to
    the atomic transition $|r\rangle \to |e\rangle$. The dynamics of storage is
    tailored by optimizing the functional dependence of the laser amplitude on
    time, $\Omega(t)$: Ideally, the atom undergoes a Raman transition to the
    final state $|r\rangle$ and the photon is stored. We analyse the storage
    \red{efficiency} including the spontaneous decay with rate $\gamma$ of the
    excited state and photon absorption or scattering at the cavity mirrors via
    an incoherent process at rate $\kbad$. Further parameters are defined in
    the text.}
\end{figure}

The purpose of this work is to provide a systematic theoretical analysis of the
efficiency of protocols for a quantum memory for single photons, where
information is stored in the electronic excitation of a single atom inside a
high-finesse resonator. \red{The qubit can be the photon
  polarization~\cite{Reiserer2015,Fleischhauer2000}, or a time-bin
  superposition of photonic states~\cite{Dilley2012}, and shall then be
  transferred into a superposition of atomic spin states.}

The scheme is illustrated in Fig. \ref{Fig:1}: a photon propagating along a
transmission line impinges on the cavity mirror, the storage protocol
coherently transfers the photon into a metastable atomic state, here denoted by
$\ket{r}$, with the help of an external laser. The protocols we analyse are
based on the seminal proposal by Cirac et al.~\cite{Cirac1997}. \red{Here, we
  first compare adiabatic protocols, originally developed for atomic ensembles
  in bad cavities~\cite{Fleischhauer2000,Gorshkov2007a} as well as a protocol
  developed for any coupling regime for a single atom~\cite{Dilley2012}. We
  then extend the protocol of Ref.~\cite{Gorshkov2007a} to quantum memories
  composed of single atoms confined inside a high-finesse resonator.} We
investigate how the storage \red{efficiency} is affected by parasitic losses at
the cavity mirrors and whether these effects can be compensated by the dynamics
induced by the laser pulse driving the atom. We finally extend our study to the
non-adiabatic regime, and analyse the \red{efficiency} of storage of broadband
photon pulses using optimal control.

This manuscript is organized as follows. In Sec. \ref{sec:model} we introduce
the basic model, which we use in order to determine the \red{efficiency} of the
storage process. In Sec. \ref{sec:adiabatic} we analyse the efficiency of
protocols based on adiabatic dynamics in presence of irreversible cavity
losses. In Sec. \ref{sec:oct} we investigate the storage efficiency when the
photon coherence time does not fulfil the condition for adiabatic quantum
dynamics. Here, we use optimal control theory to determine the shortest photon
pulse that can be stored. The conclusions are drawn in Sec.
\ref{sec:conclusions}. The appendices provide further details of the analyses
presented in Sec. \ref{sec:adiabatic}.

\section{\label{sec:model}Basic model}
The basic elements of the dynamics are illustrated in Fig. \ref{Fig:1}. A
photon propagates along the transmission line and impinges on the mirror of a
high-finesse cavity. Here, it interacts with a cavity mode at frequency $\wc$.
The cavity mode, in turn, couples to a dipolar transition of a single atom,
which is confined within the resonator. We denote by $\ket g$ the initial
electronic state in which the atom is prepared, it is a metastable state and it
performs a transition to the excited state $\ket e$ by absorbing a cavity
photon. The relevant atomic levels are shown in subplot (b): they are two
meta-stable states, $|g\rangle$ and $|r\rangle$, which are coupled by electric
dipole transitions to a common excited state $|e\rangle$ forming a $\Lambda$
level scheme. Transition $\ket{r}\to\ket{e}$ is driven by a laser, which we
model by a classical field.

In order to describe the dynamics of the photon impinging onto the cavity
mirror we resort to a coherent description of the modes of the electromagnetic
field outside the resonator. The incident photon is an excitation of the
external modes, and it couples with the single mode of a high-finesse resonator
via the finite transmittivity of the mirror on which the photon is incident.

In this section we provide the details of our theoretical model and introduce
the physical quantities which are relevant to the discussions in the rest of
this paper.

\subsection{Master equation}
The state of the system, composed of the cavity mode, the atom, and the modes
of the transmission line, is described by the density operator $\hat{\rho}$.
Its dynamics is governed by the master equation ($\hbar=1$)
\begin{equation}
  \label{eq:mastereq}
  \partial_t\hat{\rho} = -\I\commutator{\hat{H}(t)}{\hat{\rho}}
  +\Lio_\mathrm{dis}\hat{\rho}\,,
\end{equation}
where Hamiltonian $\hat{H}(t)$ describes the coherent dynamics of the modes of
the electromagnetic field outside the resonator, of the single-mode cavity, of
the atom's internal degrees of freedom, and of their mutual coupling. The
incoherent dynamics, in turn, is given by superoperator $\Lio_\mathrm{dis}$,
and includes spontaneous decay of the atomic excited state, at rate $\gamma$,
and cavity losses due to the finite transmittivity of the second cavity mirror
as well as due to scattering and/or finite absorption of radiation at the
mirror surfaces, at rate $\kbad$.

We first provide the details of the Hamiltonian. This is composed of two terms,
$\hat{H}(t)= \hat{H}_\mathrm{fields}+\hat{H}_\mathrm{I}(t)$. The first term,
$\hat{H}_\mathrm{fields}$, describes the coherent dynamics of the fields in
absence of the atom. It reads
\begin{equation}
  \label{eq:hf}
  \hat{H}_\mathrm{fields}=\sum_k(\omega_k-\wc)\hat{b}_k^\dag \hat{b}_k
  +\sum_k\lambda_k(\hat{a}^\dag \hat{b}_k+\hat{b}_k^\dag \hat{a}),
\end{equation}
and is reported in the reference frame of the cavity mode frequency $\wc$.
Here, operators $\hat{b}_k$ and $\hat{b}_k^\dag$ annihilate and create,
respectively, a photon at frequency $\omega_k$ in the transmission line, with
$\commutator{\hat{b}_k}{\hat{b}_{k'}^\dag} = \delta_{k,k'}$. The modes
$\hat{b}_k$ are formally obtained by quantizing the electromagnetic field in
the transmission line and have the same polarization as the cavity mode. They
couple with strength $\lambda_k$ to the cavity mode, which is described by a
harmonic oscillator with annihilation and creation operators $a$ and $a^\dag$,
where $ \commutator{\hat{a}}{\hat{a}^\dag} = 1$ and
$\commutator{\hat{a}}{\hat{b}_k} = \commutator{\hat{a}}{\hat{b}_k^\dag} = 0$.
In the rotating-wave approximation the interaction is of beam-splitter type and
conserves the total number of excitations. The coupling $\lambda_k$ is related
to the radiative damping rate of the cavity mode by the rate $\kappa =
L|\lambda(\wc)|^2/c$, with $\lambda(\wc)$ the coupling strength at the
cavity-mode resonance frequency~\cite{Carmichael} \red{and $L$ the length of
  the transmission line.} \red{Note that $\kappa$ is the cavity decay rate
  because of transmission into the transmission line and is necessary for the
  storage, while $\kbad$ is the decay rate into other modes and is only
  detrimental.}

The atom-photon interaction is treated in the dipole and rotating-wave
approximation. The transition $|g\rangle\to |e\rangle$ couples with the cavity
mode with strength (vacuum Rabi frequency) $g$. Transition $|r\rangle\to
|e\rangle$ is driven by a classical laser with time-dependent Rabi frequency
$\Omega(t)$, which is the function to be optimized in order to maximize the
probability of transferring the excitation into state $\ket{r}$. The
corresponding Hamiltonian reads
\begin{equation}
  \label{eq:hat}
  \hat{H}_\mathrm{I}=\delta\ketbra{r}{r}-\Delta\ketbra{e}{e}+
  \left[g\ketbra{e}{g}\hat{a}+\Omega(t)\ketbra{e}{r}+\mathrm{H.c.}\right],
\end{equation}
where $\Delta = \wc - \omega_e$ is the detuning between the cavity frequency
$\wc$ and the frequency $\omega_e$ of the $\ket{g}-\ket{e}$ transition, while
$\delta = \omega_r+\wl-\wc$ is the two-photon detuning which is evaluated using
the central frequency $\wl$ of the driving field $\Omega(t)$. \red{Here, we
  denote by $\omega_r=(E_r-E_g)/\hbar$ the frequency difference (Bohr
  frequency) between the state $\ket{r}$ (of energy $E_r$) and the state
  $\ket{g}$ (of energy $E_g$).} Unless otherwise stated, in the following we
assume that the condition of two-photon resonance $\delta=0$ is fulfilled.

The irreversible processes that we consider in our theoretical description are
(i) the radiative decay at rate $\gamma$ from the excited state $\ket{e}$,
where photons are emitted into free field modes other than the modes
$\hat{b}_k$ introduced in Eq.~\eqref{eq:hf}, and (ii) the cavity losses at rate
$\kbad$ due to absorption and scattering at the cavity mirrors and to the
finite transmittivity of the second mirror. We model each of these phenomena by
Born-Markov processes described by the superoperators $\Lio_\gamma$ and
$\Lio_{\kbad}$, respectively, such that $ \Lio_\mathrm{dis} = \Lio_\gamma +
\Lio_{\kbad}$ and
\begin{subequations}\label{eq:liouvillian}
  \begin{gather}
    \label{eq:liouvilliangamma}
    \mathcal{L}_\gamma\hat{\rho} = \gamma(2\ketbra{\xi_e}{e}\hat{\rho}
    \ketbra{e}{\xi_e}-\ketbra{e}{e}\hat{\rho}-\hat{\rho}\ketbra{e}{e})\,, \\
    \label{eq:liouvilliankappabad}
    \mathcal{L}_\kbad\hat{\rho} = \kbad(2\hat{a}\hat{\rho} \hat{a}^\dag -
    \hat{a}^\dag \hat{a}\hat{\rho}-\hat{\rho} \hat{a}^\dag \hat{a})\,.
  \end{gather}
\end{subequations}
Here, $\ket{\xi_e}$ is an auxiliary atomic state where the losses of atomic
population from the excited state $\ket{e}$ are collected.

\subsection{\label{sec:state-description}Initial state and target state}
The model is one dimensional, the transmission line is at $x<0$, and the cavity
mirror is at position $x=0$. The single incident photon is described by a
superposition of single excitations of the modes of the external
field~\cite{Blum2013}
\begin{equation}
  \label{eq:singlephoton}
  \ket{\psi_\mathrm{sp}}=\sum_k \E_k\hat{b}_k^\dag\ket{\mathrm{vac}},
\end{equation}
where $\ket{\mathrm{vac}}$ is the vacuum state and the amplitudes $\E_k$ fulfil
the normalization condition $\sum_k\abs{\E_k}^2=1$. For the studies performed
in this work, we will consider the amplitudes
\begin{equation}
  \E_k=\sqrt{\frac{c}{2L}}\int_{-\infty}^{\infty}{\rm d}t {\rm e}^{\I(kc -\wc)t}  \Ein(t)
\end{equation}
with $c$ the speed of light, $L$ the length of the transmission line, and
\begin{equation}
  \label{eq:Einhypsec}
  \Ein(t)=\frac{1}{\sqrt{T}}\sech{\br{\frac{2t}{T}}}
\end{equation}
the input amplitude at the position $x=0$, with $T$ the characteristic time
determining the coherence time $\Tc$ of the photon, $T_c=\pi T/4\sqrt{3}$ (see
definition in Eq.~(\ref{eq:Tc})). Our formalism applies to a generic input
envelope, nevertheless the specific choice of Eq. \eqref{eq:Einhypsec} allows
us to compare our results with previous studies, see
Refs.~\cite{Fleischhauer2000,Gorshkov2007a,Dilley2012}. The total state of the
system at the initial time $t=t_1$ is given by the input photon in the
transmission line, the empty resonator, and the atom in state $\ket{g}$. In
particular, the dynamics is analysed in the interval $t\in[t_1,t_2]$, with
$t_1<0$, $t_2>0$ and $|t_1|, t_2\gg \Tc$, such that (i) at the initial time
there is no spatial overlap between the single photon and the cavity mirror and
(ii) assuming that the cavity mirror is perfectly reflecting, at $t=t_2$ the
photon has been reflected away from the mirror.

The initial state is described by the density operator
$\rho(t_0)=\ket{\psi_0}\langle \psi_0|$, where
\begin{equation}
  \ket{\psi_0}=\ket{g}\otimes\ket{0}_c\otimes\ket{\psi_\mathrm{sp}}\,,
\end{equation}
and $\ket{0}_c$ is the Fock state of the resonator with zero photons.

Our target is to store the single photon into the atomic state $\ket{r}$ by
shaping the laser field $\Omega(t)$. When comparing different storage
approaches, it is essential to have a figure of merit characterizing the
performance of the process. In accordance with Ref.~\cite{Gorshkov2007a} we
define the \red{efficiency} $\eta$ of the process as the ratio between the
probability to find the excitation in the state $\ket{\psi_T}=\ket{r}\otimes
\ket{0}_c\otimes\ket{\mathrm{vac}}$ at time $t$ and the number of impinging
photons between $t_1$ and $t$, namely
\begin{equation}
  \label{eq:eta}
  \eta(t) = \frac{\langle \psi_T|\rho(t)\ket{\psi_T}}{\int_{t_1}^{t}\abs{\Ein(t)}^2\de t}\,,
\end{equation}
where $t>t_1$ and the denominator is unity for $t\to t_2$. We note that states
$\ket{\psi_0}$ and $\ket{\psi_T}$ are connected by the coherent dynamics via
the intermediate states $\ket{e}\otimes \ket{0}_c \otimes\ket{\mathrm{vac}}$
and $\ket{g} \otimes\ket{1}_c\otimes \ket{\mathrm{vac}}$. These states are
unstable, since they can decay via spontaneous emission or via the parasitic
cavity losses. Moreover, the incident photon can be reflected \red{off the
  cavity}. The latter is a unitary process, which results in a finite
probability of finding a photon excitation in the transmission line after the
photon has reached the mirror. The choice of $\Omega(t)$ shall maximize the
transfer $\ket{\psi_0}\to\ket{\psi_T}$ by minimizing the losses as well as
reflection at the cavity mirror.

\subsection{Relevant quantities}
The transmission line is here modelled by a cavity of length $L$, with a
perfect mirror at $x=-L$. The second mirror at $x=0$ coincides with the mirror
of finite transmittivity, separating the transmission line from the optical
cavity. The length $L$ is chosen to be sufficiently large to simulate a
continuum of modes for all practical purposes. This requires that the distance
between neighbouring frequencies is smaller than all characteristic frequencies
of the problem. The smallest characteristic frequency is the bandwidth of the
incident photon, which is the inverse of the \red{photon duration in time.
  Since the initial state is assumed to be a single photon in a pure state, the
  latter coincides with the photon coherence time $T_c$~\cite{Muller2017} which
  is defined as}
\begin{equation}
  \label{eq:Tc} 
  \Tc = \sqrt{\langle t^2\rangle - \langle t \rangle^2} 
\end{equation}
with $\langle t^x \rangle \equiv \int_{t_1}^{t_2}t^x\abs{\Ein(t)}^2\de t$, and
\begin{equation}
  \label{eq:einnorm}
  \int_{t_1}^{t_2}\abs{\Ein(t)}^2\de t = 1-\varepsilon\,,
\end{equation}
where $\varepsilon<10^{-5}$ for the choice $|t_1|=t_2=6\Tc$ and $L=12c\Tc$. The
modes of the transmission line are standing waves with wave vector along the
$x$ axis. For numerical purposes we take a finite number $N$ of modes around
the cavity wave number $k_\mathrm{c}=\frac{\wc}{c}$. Their wave numbers are
\begin{equation}
  \label{eq:ffmodes}
  k_n =  k_\mathrm{c} + \frac{n\pi}{L}\,,
\end{equation}
while $n=-(N-1)/2,\dots,(N-1)/2$, and the corresponding frequencies are $
\omega_n=ck_n$. We choose $N$ and $L$ so that our simulations are not
significantly affected by the finite size of the transmission line and by the
cutoff in the mode number $N$. We further choose $N$ in order to appropriately
describe spontaneous decay by the cavity mode. This is tested by initialising
the system with no atom and one cavity photon and choosing the parameters so to
reproduce the exponential damping of the cavity field.

\red{Note that a single mode of the cavity is sufficient to describe the
  interaction with a single photon if the photon frequencies lie in a range
  which is smaller than the free spectral range of the cavity and is centered
  around the frequency of the cavity mode. In this work we choose the central
  frequency of the photon to coincide with the cavity mode frequency
  $\omega_{\mathrm{p}} = \wc$ and the spectrally broadest photon we consider
  (Figs.~\ref{fig:4} and~\ref{fig:5}) spans about $16\times2\pi\unit{MHz}$
  around the cavity frequency $\wc$. A cavity of $1\unit{cm}$ has a free
  spectral range of about $15\times2\pi\unit{GHz}$ which is three orders of
  magnitudes larger than the bandwidth of the photon. This justifies the
  approximation to a single mode cavity.} \red{The employed formalism can be
  applied to photons with other center frequencies as well, if the number of
  modes $N$ is chosen sufficiently large and their center is appropriately
  shifted (c.f. eq.~(\ref{eq:ffmodes})).}

Since the free field modes are included in the unitary evolution, it is
possible to constantly monitor their state. The photon distribution in space at
time $t$ is given by
\begin{equation}
  \label{eq:photondistrspace}
  P(x,t) = \frac{2}{L} \sum_{n,m=1}^N\rho_{nm}(t)
  \sin\br{n\frac{\pi}{L}x}\sin\br{m\frac{\pi}{L}x},
\end{equation}
where $\rho_{nm}(t)={\rm Tr}\{{\hat{\rho}(t)}\ketbra{1_m}{1_n}\}$ and
$\ket{1_n}=b_{k_n}^\dagger\ket{\rm vac}$.

A further important quantity characterizing the coupling between cavity mode
and atom is the cooperativity $C$, which reads~\cite{Gorshkov2007a}
\begin{equation}
  \label{eq:C}
  C = \frac{g^2}{\kappa\gamma}\,. 
\end{equation}
The cooperativity sets the maximum storage \red{efficiency} in the limit in
which the cavity can be adiabatically eliminated from the dynamics of the
system~\cite{Gorshkov2007a}, which corresponds to assuming the condition
\begin{equation}
  \label{eq:adiabatic}
  \gamma C \Tc \gg1\,.
\end{equation}
In this limit, in fact, the state
$\ket{g}\otimes\ket{1}_c\otimes\ket{\text{vac}}$ can be eliminated from the
dynamics. Then, the \red{efficiency} satisfies $\eta(t)\le \eta_\mathrm{max}$
where the maximal \red{efficiency} $\eta_\mathrm{max}$
reads~\cite{Gorshkov2007a}
\begin{equation}
  \label{eq:etamax}
  \eta_\mathrm{max} = \frac{C}{1+C}.
\end{equation}
The maximal \red{efficiency} $\eta_\mathrm{max}$ is reached for any input
photon envelope $\Ein(t)$ and detuning $\Delta$, provided the adiabatic
condition \eqref{eq:adiabatic} is fulfilled.

\red{In our study we also determine the probability that the photon is in the
  transmission line,
  \begin{equation}
    \label{eq:pr}
    P_r(t)=\sum_k{\rm Tr}\{\hat{\rho}(t)|1_k\rangle\langle 1_k|\},
  \end{equation} 
  the probability that spontaneous emission occurs,
  \begin{equation}
    \label{eq:ps}
    P_s(t)={\rm Tr}\{\hat{\rho}(t)|\xi_e\rangle\langle \xi_e|\},
  \end{equation} 
  and finally, the probability that cavity parasitic losses take place,
  \begin{equation}
    \label{eq:ploss}
    P_\mathrm{loss}(t)=\Tr\brrr{\hat{\rho}(t)
      \ketbra{g,0_c,\text{vac}}{g,0_c,\text{vac}}}.
  \end{equation} }
By means of these quantities we gain insight into the processes leading to
optimal storage.

\red{\section{\label{sec:adiabatic}Storage in the adiabatic regime}

  In this section we determine the efficiency of storage protocols derived in
  Refs.~\cite{Fleischhauer2000,Gorshkov2007a,Dilley2012} for the setup of
  Ref.~\cite{Koerber2017} in the adiabatic regime.} We then analyse how the
\red{efficiency} of these protocols is modified by the presence of parasitic
losses at rate $\kbad$. In this case, we find also an analytic result which
corrects the maximal value of Eq. \eqref{eq:etamax}.

\red{We remark that in Refs.~\cite{Fleischhauer2000,Gorshkov2007a,Dilley2012}
  the optimal pulses $\Omega(t)$ were analytically determined using
  input-output theory~\cite{Walls1994}. In
  Refs.~\cite{Fleischhauer2000,Gorshkov2007a} the authors consider an atomic
  ensemble inside the resonator in the adiabatic regime. This regime consists
  in assuming the bad cavity limit $\kappa\gg g$ and the limit $\gamma\Tc
  C\gg1$. The first assumption allows one to adiabatically eliminate the cavity
  field variables from the equations of motion, the second assumption permits
  one to eliminate also the excited state $\ket{e}$. In Ref.~\cite{Dilley2012}
  a single atom is considered and there is no such adiabatic approximation, but
  the coupling with the external field is treated using a phenomenological
  model.

  Here we simulate the full Hamiltonian dynamics of the external field in the
  transmission line and consider a quantum memory composed of a single atom
  inside a reasonably good cavity.} The parameters we refer to in our study are
the ones of the setup of Ref.~\cite{Koerber2017}:
\begin{equation}
  \label{eq:parameters}
  (g,\kappa,\gamma)=(4.9,2.42,3.03)\MHz,  
\end{equation}
corresponding to the cooperativity $C=3.27$ and to the maximal storage
\red{efficiency} $\eta_\mathrm{max}=0.77$. When we analyse the dependence of
the \red{efficiency} on $\gamma$ or $\kappa$, we vary the parameters around the
values given in Eq.~(\ref{eq:parameters}).

\subsection{Ideal resonator}
\red{We first review the requirements and results of the individual protocols
  of Refs.~\cite{Fleischhauer2000,Gorshkov2007a,Dilley2012} and investigate
  their efficiency for a single-atom quantum memory. The works of
  Refs.~\cite{Fleischhauer2000,Gorshkov2007a,Dilley2012} determine the form of
  the optimal pulse $\Omega(t)$ for cavities with cooperativities $C\ge 1$.}
The optimal pulse is found by imposing similar, but not equivalent
requirements. In Refs.~\cite{Fleischhauer2000,Dilley2012} the authors determine
$\Omega(t)$ by imposing impedance matching, namely, that there is no photon
reflected back by the cavity mirror. In Ref.~\cite{Gorshkov2007a} the pulse
$\Omega(t)$ warrants maximal storage, namely, maximal probability of
transferring the photon into the atomic excitation $\ket{r}$. The latter
requirement corresponds to maximizing the storage \red{efficiency} $\eta$
defined in Eq.~\eqref{eq:eta}.

In detail, in Ref.~\cite{Fleischhauer2000} the authors determine the optimal
pulse $\Omega(t)$ that suppresses back-reflection from the cavity and warrants
that the dynamics follows adiabatically the dark state of the system composed
by cavity and atom. For this purpose the authors impose that the cavity field
is resonant with the transition $\ket{g}\to \ket{e}$, namely $\Delta=0$. They
further require that the coherence time $\Tc$ is larger than the cavity decay
time, $\kappa\Tc\gtrsim1$. Under these conditions the optimal pulse $\Omega(t)=
\OF(t) $ reads
\begin{equation}
  \label{eq:OF}
  \OF(t) = \frac{g \Ein(t)}{\sqrt{c_1+2\kappa\int_{t_1}^t
      \abs{\Ein(t')}^2 \de t' -\abs{\Ein(t)}^2 }}\,,
\end{equation}
where $c_1$ regularize $\OF(t)$ for $t \to t_1$. \red{The work in
  Ref.~\cite{Dilley2012} imposes the suppression of the back-reflected photon
  without any adiabatic approximation} and finds the optimal pulse $\Omega(t)=
\OD(t) $, which takes the form
\begin{equation}
  \label{eq:OD}
  \OD(t)=\frac{g\Ein(t)+\left(\dot{\F}(t)+\gamma\F(t)\right)/g}
  {\sqrt{2\kappa\rho_0+2\kappa\int\limits_{t_1}^t
      \abs{\Ein(t')}^2-\abs{\Ein(t)}^2-\mathcal D(t)}},
\end{equation} 
with
\begin{equation}
  \mathcal D(t)=2\gamma \int\limits_{t_1}^t\abs{\F(t')}^2\de t'+\abs{\F(t)}^2\,.
\end{equation}
and $ \F(t) = \dot{\Ein}(t)-\kappa\Ein(t)$. Coefficient $\rho_0$ accounts for a
small initial population in the target state $\ket{r}$ and it is relevant in
order to avoid divergences in Eq.~\eqref{eq:OD} for $t\to t_1$, \red{see
  Ref.~\cite{Dilley2012} for an extensive discussion.} The pulse $\OF(t)$ of
Eq.~\eqref{eq:OF} can be recovered from Eq.~\eqref{eq:OD} by imposing the
conditions
\begin{subequations}
  \begin{gather}
    \label{eq:dilley2fleischhauer1}
    \dot{\F}(t)+\gamma\F(t)=0,\,   \\
    \label{eq:dilley2fleischhauer2}
    -\abs{\F(t_1)}^2 + 2\kappa\rho_0 = c_1.
  \end{gather}
\end{subequations}
The control pulse $\OD(t)$ can be considered as a generalization of $\OF(t)$
since it is determined by \red{solely} imposing quantum impedance matching.
 
In Ref.~\cite{Gorshkov2007a} the authors determine the amplitude $\Omega(t)$
that maximizes the \red{efficiency} $\eta$. This condition is not equivalent to
imposing impedance matching. In fact, while in the case of impedance matching
major losses through the excited state $\ket{e}$ are acceptable in order to
minimize the probability of photon reflection, in the case of maximum transfer
\red{efficiency} $\eta$ those losses are detrimental and thus have to be
minimized. The optimal pulse $\Omega(t)= \OG(t)$ is determined for a generic
detuning $\Delta$ by using an analytical model based on the adiabatic
elimination of the excited state $\ket{e}$ of the atom and of the cavity field
in the bad cavity limit $\kappa\gg g$. It reads
\begin{equation}
  \label{eq:OG}
  \begin{aligned}
    \OG(t) &{}=\frac{\gamma(1+C)+\I\Delta}{\sqrt{2\gamma(1+C)}}
    \frac{\Ein(t)}{\sqrt{\int_{t_1}^t\abs{\Ein(t')}^2\de t'}}\\
    &\times\exp\br{{-\I\frac{\Delta}{2\gamma(1+C)}\ln{\int_{t_1}^{t}
          \abs{\Ein(t')}^2\de t'}}}\,.
  \end{aligned}
\end{equation}
In the limit in which the adiabatic conditions are fulfilled, this control
pulse allows for storage with \red{efficiency} $\eta_\mathrm{max}$,
Eq.~(\ref{eq:etamax}). This \red{efficiency} approaches unity for
cooperativities $C\gg 1$.

We start by integrating numerically the master equation for a single
atom~\eqref{eq:mastereq} after setting $\kbad=0$, namely, by neglecting
parasitic losses. We determine the storage \red{efficiency} at the time $t_2$,
which we identify by taking $t_2\gg\Tc$ for different choices of the control
field $\Omega=\OG,\OF,\OD$ in Hamiltonian \eqref{eq:hat}. Numerically, $t_2$
corresponds to the time the photon would need to be reflected back into the
initial position, assuming that the partially reflecting mirror is replaced by
a perfect mirror. Our numerical simulations are performed for a single atom in
a resonator in the good cavity limit.

\begin{figure*}[!htb]
  \includegraphics[]{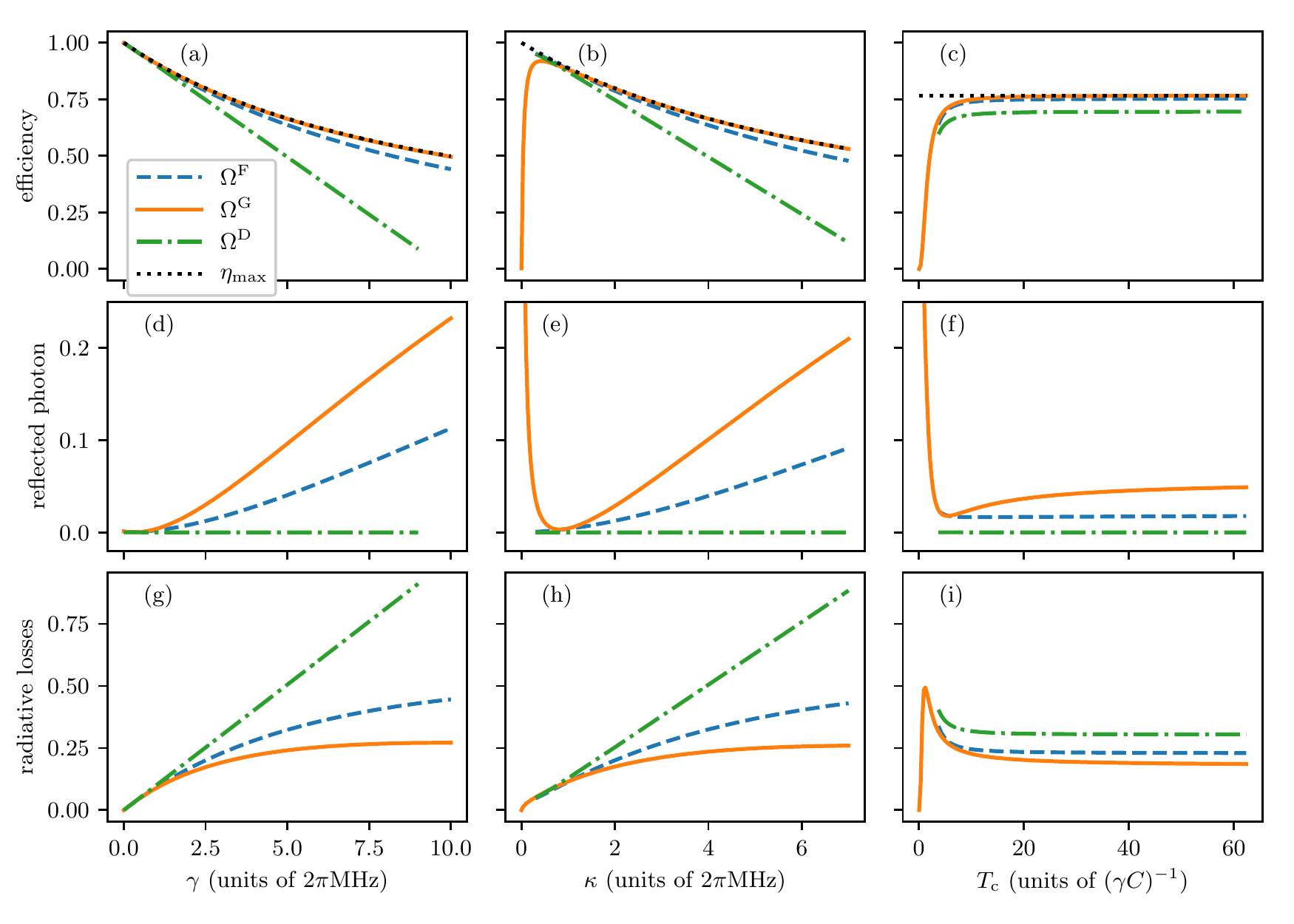} \red{
    \caption{ \label{fig:2}Comparison between the protocols of
      Refs.~\cite{Fleischhauer2000,Gorshkov2007a,Dilley2012}. (Upper row)
      Storage efficiency $\eta$, Eq.~\eqref{eq:eta}, (central row) probability
      that the photon is reflected $P_r$, Eq.~(\ref{eq:pr}), and (bottom row)
      probability of spontaneous emission, Eq.~(\ref{eq:ps}), evaluated at time
      $t_2=6\Tc$ by integrating numerically Eq.~\eqref{eq:mastereq} for
      $\kbad=0$. The quantities are reported as a function of (left column) the
      decay rate $\gamma$ from the excited state (for $\kappa=\kappa_0$ and
      $\Tc=\Tc^0$), (central column) the decay rate $\kappa$ of the cavity
      field (for $\gamma=\gamma_0$ and $\Tc=\Tc^0$) and (right column) the
      coherence time of the photon $\Tc$ (in units of $1/(\gamma C)$ and for
      $\kappa=\kappa_0$ and $\gamma=\gamma_0$). The three different lines
      $\OF$, $\OD$, and $\OG$ refer to the evolution with the respective
      control pulse (see Eqs.~\eqref{eq:OF},~\eqref{eq:OD},~\eqref{eq:OG},
      respectively). The dotted lines in panels (a)(b)(c) correspond to the
      maximum efficiency $\eta=C/(1+C)$, Eq. \eqref{eq:etamax}. Here,
      $(g,\kappa_0,\gamma_0)=(4.9,2.42,3.03)\MHz$ and $\Tc^0=0.5\unit{\mu s}$.
      The input pulse $\Ein(t)$ is given in Eq.~\eqref{eq:Einhypsec}, at the
      initial time $t_1=-6\Tc$ the pulse has negligible overlap with the cavity
      mode. The transmission line has length $L=\max(12 c\Tc,15c/\kappa)$ and
      $211$ equispaced modes. With this choice the frequency range of the modes
      included in the simulation is about $40\kappa$ around the cavity
      frequency $\wc$.}
  }
\end{figure*}

\red{Figures~\ref{fig:2} display the efficiency and the losses as a function of
  $\kappa$, $\gamma$, and of the coherence time $T_c$ of the photon (and thus
  of the adiabatic parameter $\gamma\Tc C$). Each curve corresponds to the
  different control pulses in the Hamiltonian~\eqref{eq:hat} according to the
  three protocols. In subplot~(a) we observe that the efficiency reached with
  the pulse $\OG(t)$ corresponds to the maximum theoretical efficiency
  $\eta_\mathrm{max}$, while the efficiency with $\OD$ is the smallest. In
  subplot~(b) it is visible that the control pulse $\OG(t)$ warrants the
  maximum efficiency even down to values of $\kappa$ of the order of
  $\kappa\sim g/5$. Subplot~(c) displays the efficiency as a function of the
  adiabatic parameter $\gamma\Tc C$: the protocol $\OG(t)$ reaches the maximum
  theoretical efficiency $\eta_\mathrm{max}$ for $\gamma\Tc C\gtrsim 20$, while
  the other protocols have smaller efficiency for all values of $\Tc$.
  Figures~\ref{fig:2}(d)(e)(f) report the probability that the photon is
  reflected back into the transmission line, eq.~(\ref{eq:pr}). It is evident
  that protocol $\OD$ perfectly suppresses the back reflection probability in
  every regime here considered. However in the non-adiabatic regime (subplots
  (c)(f)(i), $\gamma\Tc C \lesssim20$) the protocol $\OD$, as well as the
  protocol $\OF$, requires an increasing maximum Rabi frequency for decreasing
  $\Tc$. At the value of about $\gamma\Tc C\approx3.74$ the Rabi frequency is
  so high that it is not anymore manageable by our numerical solver, for this
  reason the plots for the protocols $\OD$ and $\OF$ are reported for
  $\gamma\Tc C\gtrsim 3.74$. The same happens for small values of $\kappa$,
  subplots (b)(e)(h): in this case the plots for the protocols $\OD$ and $\OF$
  are reported for $\kappa\gtrsim0.3\times2\pi\unit{MHz}$. The diverging Rabi
  frequency can be avoided by an appropriate choice of the parameters $c_1$ and
  $\rho_0$ in Eqs.~(\ref{eq:OF}) and~(\ref{eq:OD}), respectively.
  Figures~\ref{fig:2}(g)(h)(i) report the losses via spontaneous emission of
  the atom, Eq.~(\ref{eq:ps}): while these losses are acceptable in order to
  minimize the back-reflected photon, they are detrimental for the intent of
  populating the target state $\ket{r}$. Protocol $\OD$, which perfectly
  suppresses the back reflected photon, has the highest losses via spontaneous
  emission, which in the end leads to a lower efficiency $\eta$. Protocol $\OG$
  in turn, has the lowest radiative losses and it allows for the transfer with
  the maximal efficiency $\eta_{\mathrm{max}}$. Protocol $\OF$ tries to
  minimize reflection of the photon at the cavity mirror. However, since $\OF$
  is derived with some approximations, it does not suppress completely the
  reflection and its final efficiency is between the ones of the other two
  protocols.}

An important general result of this study is that the bad cavity limit is not
essential for reaching the maximal \red{efficiency} as long as the dynamics is
adiabatic: the relevant parameter is in fact the cooperativity.
 
\subsection{\label{sec:kbad} Parasitic losses}
The protocols so far discussed assume an ideal optical resonator. In this
section we analyse how their \red{efficiency} is modified by the presence of
parasitic losses, here described by the superoperator $\Lio_{\kbad}$ in
Eq.~(\ref{eq:liouvilliankappabad}). In particular, we derive the maximal
\red{efficiency} the protocols can reach as a function of $\kbad>0$.

\begin{figure}[!ht]
  \centering \includegraphics[]{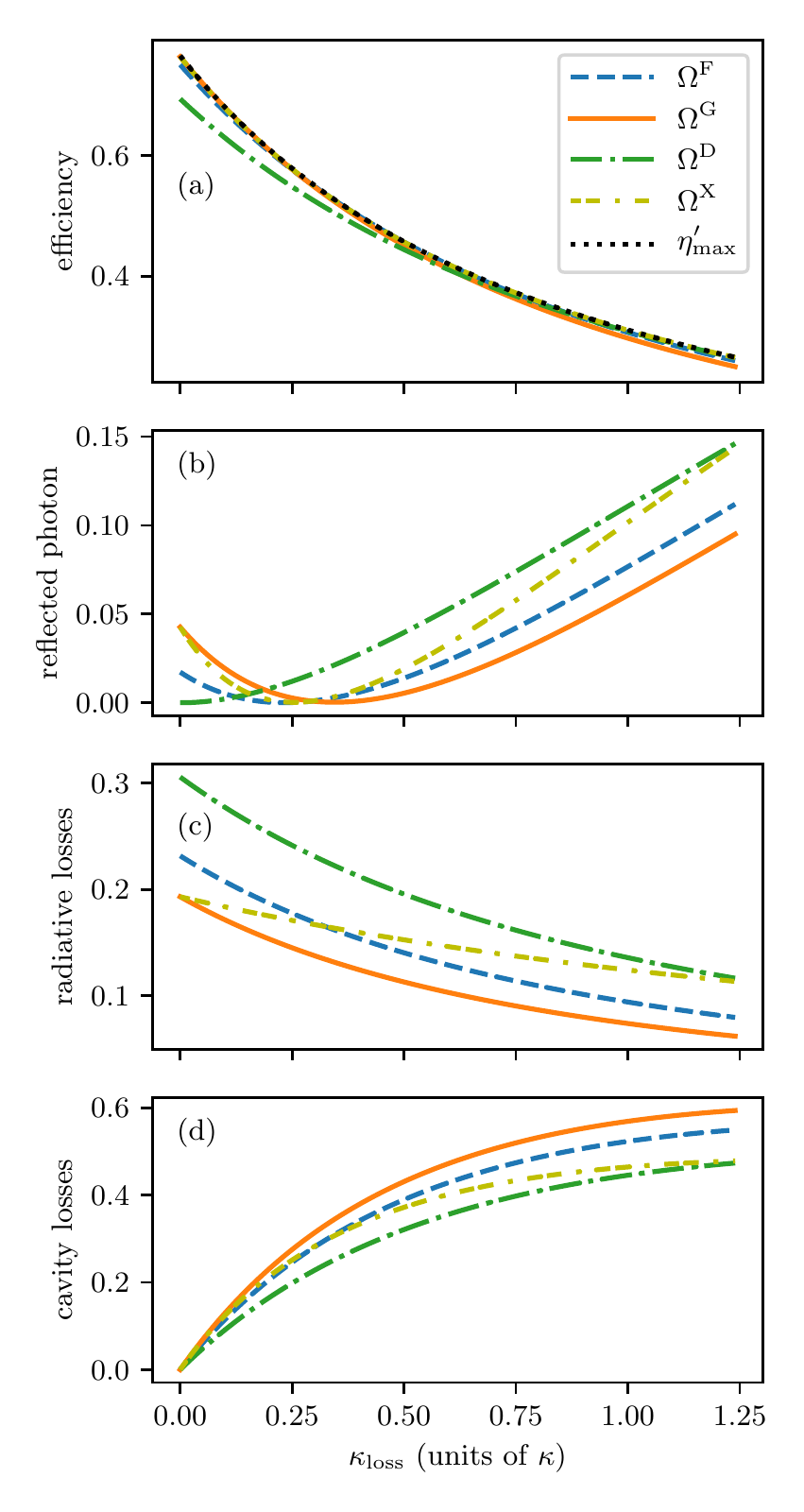} \red{
    \caption{\label{fig:3}Efficiency of storage protocols in the adiabatic
      regime as a function of the rate of parasitic losses $\kbad$ (in units of
      $\kappa$). (a) Storage efficiency, Eq. \eqref{eq:eta}, (b) the
      probability that the photon is reflected, Eq. \eqref{eq:pr}, (c) the
      probability of spontaneous decay, Eq. \eqref{eq:ps}, and (d) the
      probability of parasitic losses, Eq. \eqref{eq:ploss}, evaluated at time
      $t_2=6\Tc$ and for $(g,\kappa,\gamma)=(4.9,2.42,3.03)\MHz$,
      $\Tc=0.5\unit{\mu s}$. The other parameters are the same as in
      Fig.~\ref{fig:2}. The lines $\OX$, $\OF$, $\OD$, and $\OG$ refer to the
      evolution with the respective control pulse (resp.
      Eqs.~\eqref{eq:OX},~\eqref{eq:OF},~\eqref{eq:OD},~\eqref{eq:OG}). The
      dotted line in (a) corresponds to the value of $\eta'_\mathrm{max}$, Eq.
      \eqref{eq:etamaxNew}.}}
\end{figure}

We first numerically determine the \red{efficiency} of the individual protocols
as a function of $\kbad$ for $\Tc=0.5\unit{\mu s}$. Figure~\ref{fig:3}(a)
displays $\eta$ for $\Omega=\OG, \OD, \OF$. It is evident that the effect of
losses is detrimental, for instance it leads to a definite reduction of the
maximal \red{efficiency} from $\eta=0.77$ down to $\eta=0.68$ for $\kbad\sim
0.1\kappa$. This result can be improved by identifying a control field
$\Omega=\OX$ which compensates, at least partially, the effects of these
parasitic losses. The control field $\OX(t)$ is derived \red{in
  Sec.~\ref{sec:maxim-eff-kbad}} using the input-output formalism: it
corresponds to performing the substitution $\kappa\to \kappa + \kbad$ in the
functional form $\OG(t)$ of Eq. \eqref{eq:OG}. Specifically, it reads
\begin{eqnarray}
  \label{eq:OX}
  \OX(t) &=&\frac{\gamma(1+C')+\I\Delta}{\sqrt{2\gamma(1+C')}}
             \frac{\Ein(t)}{\sqrt{\int_{t_1}^t\abs{\Ein(t')}^2\de t'}}\\
         & &\times\exp\br{{-\I\frac{\Delta}{2\gamma(1+C')}\ln{\int_{t_1}^{t}
             \abs{\Ein(t')}^2\de t'}}}\,,\nonumber
\end{eqnarray}
with the modified cooperativity
\begin{equation}
  \label{eq:newcooperativity}
  C'=\frac{g^2}{\gamma(\kappa+\kbad) }\,.
\end{equation}
When the control pulse $\OX(t)$ is used, the \red{efficiency} of the process
corresponds to the maximal \red{efficiency} $\eta'_\mathrm{max}$, which is now
given by
\begin{equation}
  \label{eq:etamaxNew}
  \eta'_\mathrm{max} =\frac{\kappa}{\kappa+\kbad}\frac{C'}{1+C'}\,.
\end{equation}
Clearly, $\eta'_\mathrm{max}\le\eta_\mathrm{max}$, while the equality holds for
$\kbad=0$.

By inspecting the numerical results, we note that the \red{efficiency} obtained
using $\OX$ is always higher than the one reached by the other protocols. Even
though for some values of $\kbad$ the \red{efficiencies} using different
control fields may approach the one found with $\OX$, yet the dynamics are
substantially different. This is visible by inspecting the probability that the
photon is reflected, the radiative losses, and the parasitic losses, as a
function of $\kbad$ as shown in Figs.~\ref{fig:3}(b)(c)(d), respectively: Each
pulse distributes the losses in a different way, with $\OX(t)$ interpolating
among the different strategies in order to maximize the \red{efficiency}.

\begin{figure}[!ht]
  \centering \includegraphics[]{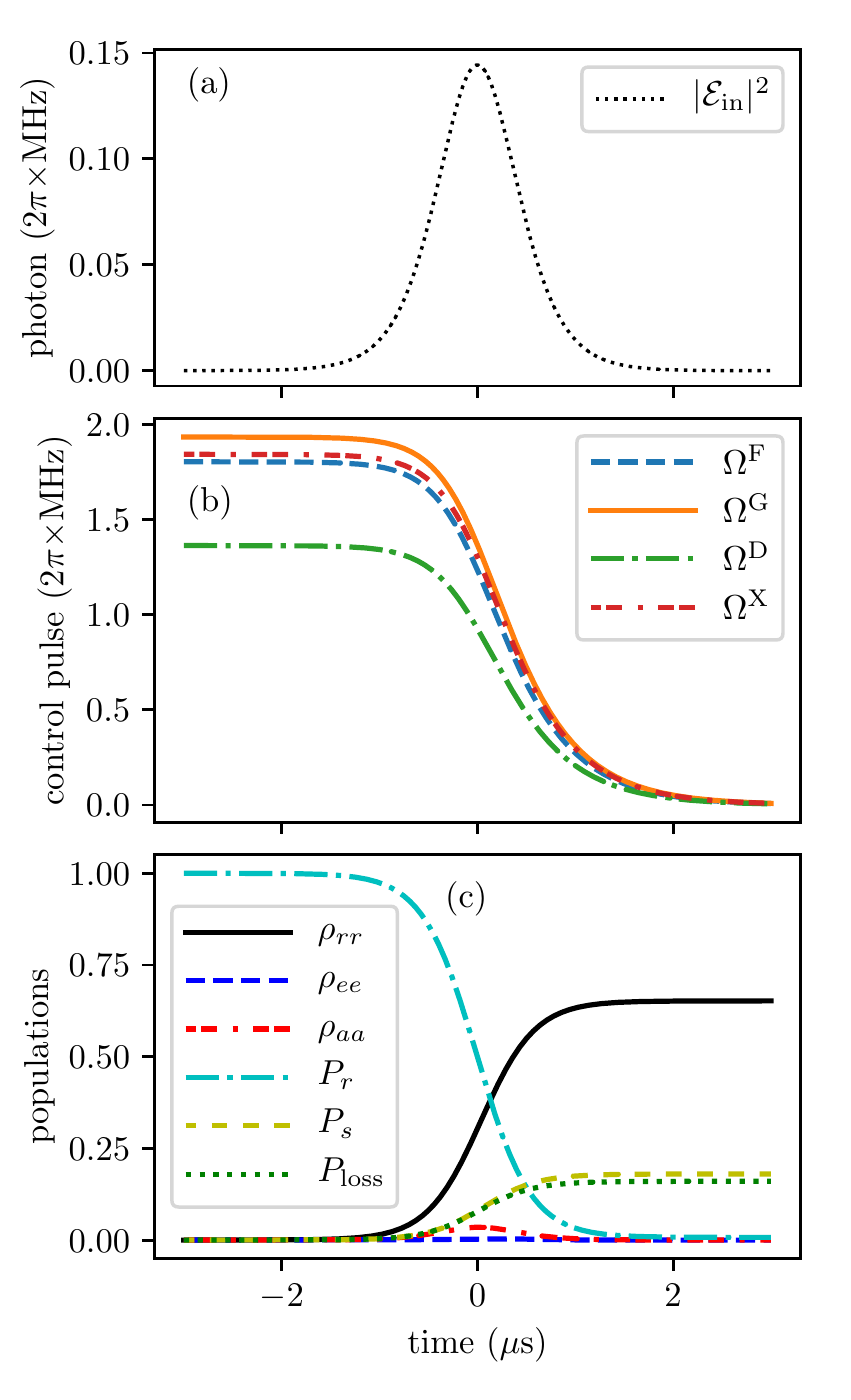} \red{
    \caption{\label{fig:pulse_evolution} Dynamics of storage. (a) Photon
      envelope $\abs{\Ein(t)}^2$, Eq.~(\ref{eq:Einhypsec}), as a function of
      time. (b) Time dependence of the control pulses $\OF(t)$, $\OG(t)$,
      $\OD(t)$, and $\OX(t)$ (resp.
      Eqs.~\eqref{eq:OF},~\eqref{eq:OG},~\eqref{eq:OD},~\eqref{eq:OX}). (c)
      Time evolution of the diagonal elements of the density matrix when the
      atom is driven by $\OX$. The curves are the population $\rho_{rr}$ of
      state $\ket{r}$, the population $\rho_{ee}$ of state $\ket{e}$, the
      probability that there is one photon in the cavity $\rho_{aa}$, the
      probability that the photon is in the transmission line $P_r$,
      Eq.~(\ref{eq:pr}), the probability of spontaneous decay, Eq.
      \eqref{eq:ps} $P_s$, and the probability of cavity parasitic losses
      $P_{\mathrm{loss}}$, Eq. \eqref{eq:ploss}. The parameters are
      $(g,\kappa,\gamma,\kbad)=(4.9,2.42,3.03,0.33)\MHz$, $\Delta=0$ and
      $\Tc=0.5\unit{\mu s}$.} }
\end{figure}
\red{Figure~\ref{fig:pulse_evolution} shows the evolution of the system for
  $\Tc=0.5\unit{\mu s}$. Fig.~\ref{fig:pulse_evolution}(a) displays the
  envelope in time $\abs{\Ein(t)}^2$ for the photon given in
  eq.~(\ref{eq:Einhypsec}), which is the one used also in this simulation.
  Fig.~\ref{fig:pulse_evolution}(b) displays the control pulse shapes of the
  protocols $\OF$, $\OG$, $\OD$, of
  Refs.~\cite{Fleischhauer2000,Gorshkov2007a,Dilley2012} and $\OX$ derived in
  this work (the pulse shapes are given analytically in
  Eqs.~\eqref{eq:OF},~\eqref{eq:OG},~\eqref{eq:OD},~\eqref{eq:OX}).
  Fig.~\ref{fig:pulse_evolution}(c) shows the population of the states and the
  losses during the evolution when the atom is driven by $\OX(t)$. The
  efficiency of the transfer, Eq.~(\ref{eq:eta}), corresponds to the population
  of the state $\ket{r}$, $\rho_{rr}$. For the parameters of
  Ref.~\cite{Koerber2017} the final efficiency is
  $\eta(t_2)\approx\eta'_{\mathrm{max}}\approx0.653$.}

In the next subsection we report the derivation of $\OX$ and
$\eta'_\mathrm{max} $ by means of the input-output formalism.

\subsection{\label{sec:maxim-eff-kbad}Maximal \red{efficiency} in presence of
  parasitic losses}
In this section we generalize the adiabatic protocol of
Ref.~\cite{Gorshkov2007a} in order to identify the control field that maximizes
the storage \red{efficiency} and to determine the maximum storage
\red{efficiency} one can reach. The derivation presented in this section is
based on the input-output formalism and it delivers Eq.~\eqref{eq:OX} and
Eq.~\eqref{eq:etamaxNew}.

We first justify the result for Eq.~\eqref{eq:etamaxNew} using a time reversal
argument applied in Refs.~\cite{Gorshkov2007a,Gorshkov2007b}. Let us consider
retrieval of the photon, assuming the atom is initially in state $\ket{r}$ and
there is neither external nor cavity field. Then, in order to retrieve the
photon, the control pulse $\Omega(t)$ shall drive the transition
$\ket{r}\to\ket{e}$ such that at the end of the process the state $\ket{r}$ is
completely empty. The excited state $\ket{e}$ dissipates the excitation with
probability $1/(1+C')$, while it can emit into the cavity mode with probability
$C'/(1+C')$. When the cavity mode is populated, a fraction
$\kbad/(\kappa+\kbad)$ is lost, while the fraction $\kappa/(\kappa+\kbad)$ is
emitted via the coupling mirror into the transmission line. From this argument
one finds that the probability of retrieval is given by
Eq.~(\ref{eq:etamaxNew}). Using the time reversal argument, this is also the
\red{efficiency} of storage.

We now derive this result as well as $\OX(t)$ starting from the retrieval
process and then applying the time reversal argument. For this purpose, we
restrict the dynamics to the Hilbert space ${\mathcal H}$ composed by the
states $\{\ket{g,1_c,\text{vac}},\ket{e,0_c,\text{vac}},\ket{r,0_c,\text{vac}},
\ket{g,0_c,1_k}: 1\le k\le N\}$. In ${\mathcal H}$ the probability is not
conserved due to leakage via spontaneous decay and via parasitic cavity losses.
Therefore, a generic state in ${\mathcal H}$ takes the form $\ket{\phi(t)}
=c(t)\ket{g,1_c,\text{vac}}+e(t)\ket{e,0_c,\text{vac}}
+r(t)\ket{r,0_c,\text{vac}}+\sum_k\mathcal{E}_k(t)\ket{g,0_c,1_k}$, it evolves
according to a non-Hermitian Hamiltonian and its norm decays exponentially with
time~\cite{Moelmer1988}. We assume that at the initial time $t=t_1$ the
probability amplitude $r(t_1)$ equals $1$, while all other probability
amplitudes vanish. The equations of motion for the probability amplitudes read
\begin{subequations}
  \label{eq:phi1_inout}
  \begin{gather}
    \dot{c}(t) =-\I g e(t)-\I\sqrt{2\kappa}\Ein(t)-(\kappa+\kbad )c(t),\\
    \dot{e}(t) =(\I\Delta -\gamma )e(t)-\I g c(t) -\I\Omega(t)r(t),\\
    \dot{r}(t) = -\I\Omega^*(t)e(t)\,,
  \end{gather}
\end{subequations}
where we used the Markov approximation and the input-output formalism
\cite{Walls1994}. We now assume the bad-cavity limit $\kappa\gg g$ and
adiabatically eliminate the cavity field from the equations of motion (which
corresponds to assuming $\dot c(t)\approx 0$ over the typical time scales of
the other variables). In this limit the input-output operator relation,
$\hat{\mathcal{E}}_\mathrm{out}(t) = \I\sqrt{2\kappa}\hat{a}(t)
-\hat{\mathcal{E}}_\mathrm{in}(t)$, takes the form
\begin{eqnarray}
  \label{eq:inout_cavadeliminated}
  \Eout(t) = G\sqrt{2\gamma C}e(t)+\frac{\kappa-\kbad}{\kappa+\kbad}\Ein(t)\,,
\end{eqnarray}
where $$G=\kappa/(\kappa+\kbad)$$ and $C$ is given in Eq. \eqref{eq:C}. This
equation has to be integrated together with the equations
\begin{align}
  \label{eq:inout_cavadeliminated:2}
  &\dot{e}(t) =\brr{\I\Delta -\gamma\br{1+GC}}e(t) -\I\Omega(t)r(t)
    -G\sqrt{2\gamma C}\Ein(t)\,,\\
  &\dot{r}(t) = -\I\Omega^*(t)e(t)\,.
\end{align}
Our goal is to determine the retrieval \red{efficiency} assuming that at time
$t=0$ there is no input photonic excitation, thus $\Ein(t)=0$ at all times.
Using these assumptions, the above equations can be cast into the form
\begin{equation}
  \label{eq:step1foretaprime}
  \td{}{t}\br{\abs{e(t)}^2+\abs{r(t)}^2} = -2\gamma\br{1+C'}\abs{e(t)}^2\,.
\end{equation}
The probability that no excitations are left in the atom at time $t_2>0$
($t_2\gg\Tc$) is the retrieval \red{efficiency}
\begin{equation}
  \label{eq:retrfid}
  \begin{aligned}
    \eta'_\mathrm{max} &= \int_{t_1}^{t_2}\abs{\Eout(t)}^2\de t
    = 2G^2\gamma C\int_{t_1}^{t_2}\abs{e(t)}^2\de t =  \\
    &= \frac{-GC'}{1+C'}\brr{\abs{e(t)}^2+\abs{r(t)}^2}_{t_1}^{t_2}=
    \frac{GC'}{1+C'}\,.
  \end{aligned}
\end{equation}
By means of the time reversal argument, this is also the storage
\red{efficiency}.

The output field can be analytically determined by adiabatically eliminating
the excited state from Eqs.~(\ref{eq:inout_cavadeliminated}). This leads to the
expression
\begin{equation}
  \label{eq:Eoutanalyt}
  \begin{aligned}
    \Eout(t)={}&\I \sqrt{2\gamma GC'}\frac{\Omega(t)}{\I\Delta-\gamma(1+C')} \\
    &\times\exp\br{\int_{t_1}^t\frac{\abs{\Omega(t')}^2}{\I\Delta-\gamma(1+C')}\de
      t'}\,.
  \end{aligned}
\end{equation}
Integrating the norm squared of Eq.~(\ref{eq:Eoutanalyt}) one obtains
\begin{equation}
  \label{eq:Eoputanalintsquared}
  \begin{gathered}
    \br{ G\frac{C'}{1+C'}}^{-1}\int_{t_1}^t\abs{\Eout(t')}^2\de t' = \\
    = 1-\exp\brr{\frac{-2\gamma(1+C')}{\gamma^2(1+C')^2+\Delta^2}
      \int_{t_1}^t\abs{\Omega(t')}^2\de t'}\,.
  \end{gathered}
\end{equation}
We solve Eq.~(\ref{eq:Eoputanalintsquared}) to find $\abs{\Omega(t)}$, while
the phase of $\Omega(t)$ can be determined from Eq.~(\ref{eq:Eoutanalyt}).
Finally, we obtain the control pulse $ \OX_\mathrm{retr}(t)$ which retrieves
the photon with \red{efficiency} $\eta'_\mathrm{max}$. It reads
\begin{equation}
  \label{eq:Xretr}
  \begin{aligned}
    & \OX_\mathrm{retr}(t) =\frac{\gamma(1+C')-\I\Delta}{\sqrt{2\gamma(1+C')}}
    \frac{\Eout(t)}{\sqrt{\int_{t}^{t_2}\abs{\Eout(t')}^2\de t'}}\\
    &\times\exp\br{{\I\frac{\Delta}{2\gamma(1+C')}\ln{\int_{t}^{t_2}
          (\abs{\Eout(t')}^2/\eta'_\mathrm{max})\de t'}}}\,.
  \end{aligned}
\end{equation}
Using the time reversal argument, the control pulse $\OX(t) =
\OX^*_\mathrm{retr}(T-t)$ stores the time reversed input photon with
$\Ein(t)=\Eout^*(T-t)/\sqrt{\eta'_\mathrm{max}}$ and $T=t_2-t_1$, and it takes
the form given in Eq. \eqref{eq:OX}. This pulse has the same form as the pulse
of Eq.~(\ref{eq:OG}), where now $C$ has been replaced by $C'$ (or equivalently
$\kappa \to \kappa + \kbad$).

\subsection{Photon Retrieval}
\red{The generation of single photons with arbitrary shape of the wavepacket
  envelope in atom-cavity systems has been discussed theoretically
  in~\cite{Gorshkov2007a,Vasilev2010} and demonstrated experimentally
  in~\cite{Keller2004,Nisbet-Jones2011}.

  In Ref.~\cite{Cirac1997,Gorshkov2007b} it has been pointed out that photon
  storage and retrieval are connected by a time reversal transformation.} This
argument has profound implications. Consider for instance the pulse shape
$\Omega(t)$ which optimally stores an input photon with envelope $\Ein(t)$.
This pulse shape is the time reversal of the pulse shape
$\Omega_\mathrm{retr}(t) = \Omega^*(T-t)$ which retrieves a photon with
envelope $\Eout(t)=\Ein^*(T-t)$ (here $T=t_2-t_1$). In this case, the storage
\red{efficiency} is equal to the \red{efficiency} of retrieval and is limited
by the cooperativity through the relation in Eq. \eqref{eq:etamaxNew}. We have
numerically checked that this is fulfilled by considering adiabatic retrieval
and storage of a single photon through 5 nodes, consisting of 5 identical
cavity-atom systems. We applied $\Omega_\mathrm{retr}(t)$ for the retrieval and
the corresponding $\Omega(t)$ for the storage. Within the numerical error, we
verified that the storage \red{efficiency} of each retrieved photon remains
constant and equal to the one of the first retrieved photon.

\section{\label{sec:oct}Beyond adiabaticity}

In this section we analyse the efficiency of storage of single photon pulses in
the regime in which the adiabaticity condition Eq.~\eqref{eq:adiabatic} does
not hold. Our treatment extends to single-atom quantum memories the approach
that was applied to atomic ensemble in Refs.~\cite{Novikova2007,Gorshkov2008}
and allows us to identify the minimum coherence time scale of the photon pulse
for which a given target \red{efficiency} can be reached.

Our procedure is developed as follows. We use the von-Neumann equation,
obtained from Eq. \eqref{eq:mastereq} after setting $\gamma=\kbad=0$, and
resort to optimal control theory for identifying the control pulse
$\Omega(t)=\Oo(t)$ that maximizes the storage \red{efficiency} for
$\gamma=\kbad=0$. Specifically, we make use of the GRAPE
algorithm~\cite{Khaneja2005} implemented in the library
QuTiP~\cite{Johansson2012-13}. We then determine the storage \red{efficiency}
of the full dynamics, including spontaneous decay and cavity parasitic losses,
by numerically integrating the master equation \eqref{eq:mastereq} using the
pulse $\Oo(t)$. We show that the dynamics due to $\Oo(t)$ significantly differs
from the adiabatic dynamics, and thereby improve the \red{efficiency} for short
coherence times.

\begin{figure}[!ht]
  \centering \includegraphics[]{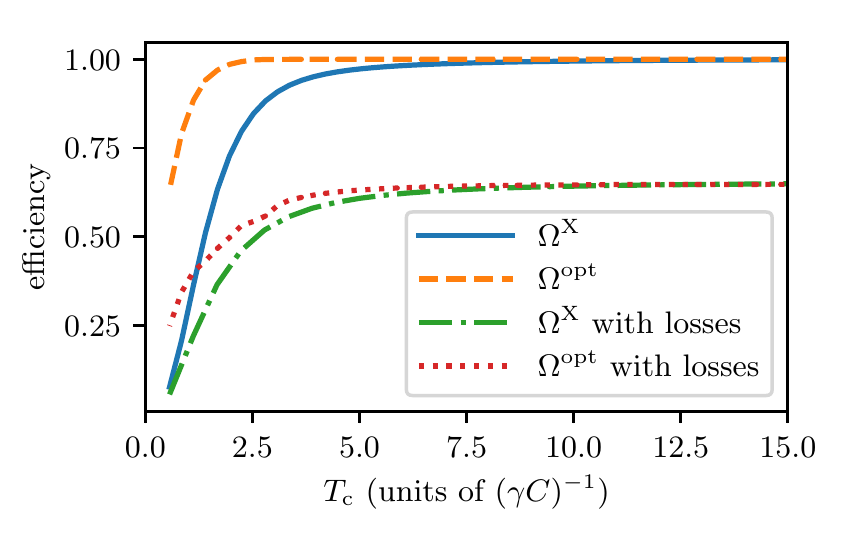}
  \caption{\label{fig:4}Storage \red{efficiency} $\eta$ at $t=t_2$ as a
    function of the coherence time of the single-photon pulse $\Tc$ (in units
    of $(\gamma C)^{-1}$). The legenda indicates the pulses used in the
    numerical integration of Eq. \eqref{eq:mastereq}. The parameters are
    $(g,\kappa)=(4.9,2.42)\MHz$, the lines labeled ``with losses'' refer to the
    \red{efficiency} of the process when $(\gamma,\kbad)=(3.03,0.33)\MHz$,
    otherwise $\gamma=\kbad=0$; $t_2=-t_1=6\Tc$. The other parameters are given
    in Fig. \ref{fig:2}.}
\end{figure}

Figure~\ref{fig:4} displays the storage \red{efficiency} $\eta$ as a function
of the photon coherence time $\Tc$ when the control pulse is $\OX(t)$, Eq.
\eqref{eq:OX}, and when instead the control pulse is found by means of the
numerical procedure specified above, which we denote by $\Oo(t)$. The storage
\red{efficiency} is reported for $\gamma=\kbad=0$ and for
$(\gamma,\kbad)=(3.03,0.33)\MHz$. The results show that optimal control, in the
way we implement it, does not improve the maximal value of the storage
\red{efficiency}, which seems to be limited by the value of $\eta_{\rm max}'$,
Eq. \eqref{eq:etamaxNew}. We remark that this behaviour is generally
encountered when applying optimal-control-based protocols to Markovian
dynamics~\cite{Koch2016}. Nevertheless, the protocols identified using optimal
control extend the range of values of $\Tc$, where the maximal \red{efficiency}
is reached, down to values where the adiabatic condition is not fulfilled. We
further find that the optimized pulse we numerically identified in absence of
losses provides an excellent guideline for optimizing the storage also in
presence of losses.

\begin{figure}[!ht]
  \centering \includegraphics[]{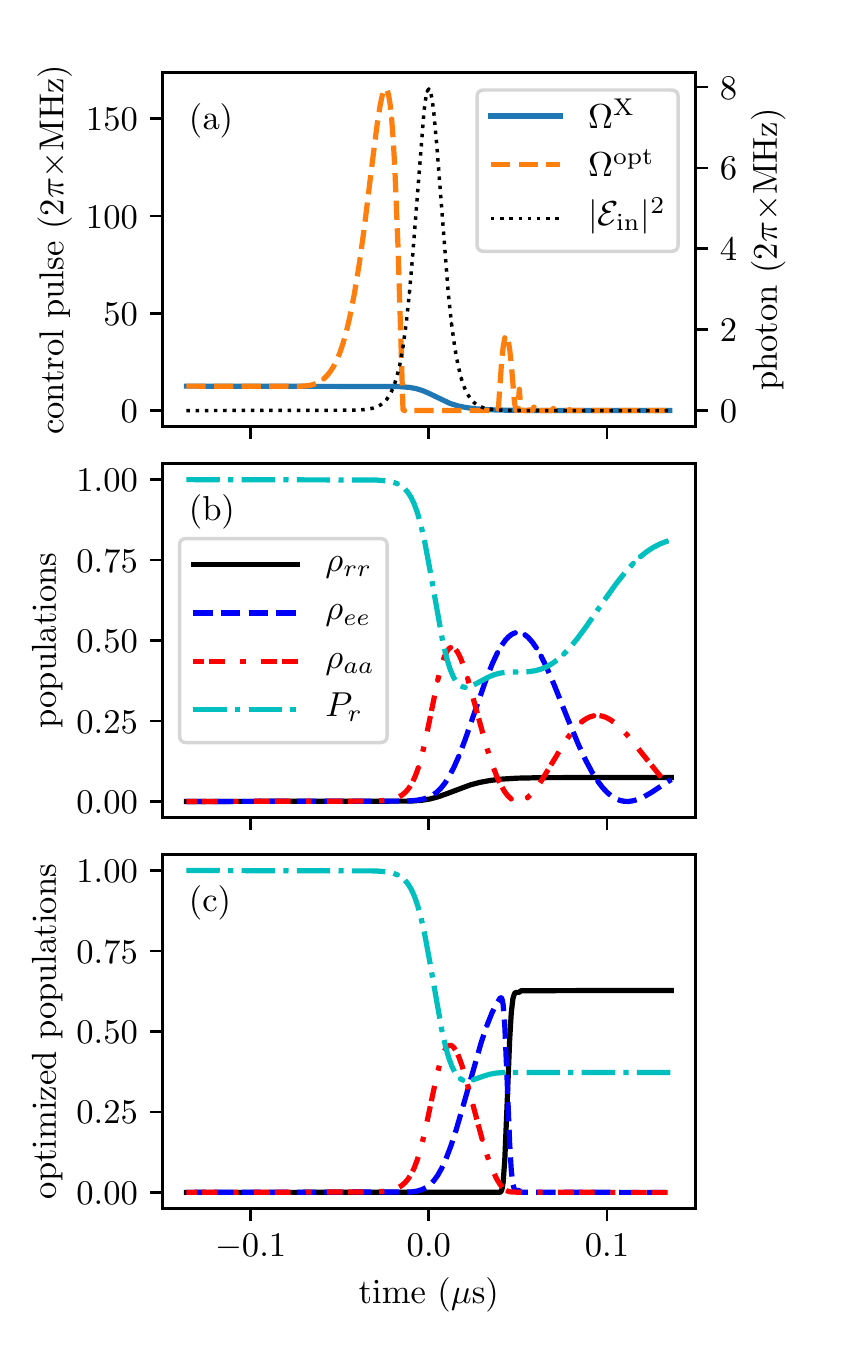}
  \red{\caption{\label{fig:5}(a) Photon envelope $\abs{\Ein(t)}^2$,
      Eq.~(\ref{eq:Einhypsec}), and optimized pulse $\Omega^\mathrm{opt}(t)$ as
      a function of time (the initial guess pulse $\OX(t)$ is shown for
      comparison). Subplot (b) and (c) display the time evolution of the
      diagonal elements of the density matrix when the atom is driven by $\OX$
      and $\Oo$, respectively. The curves are the population $\rho_{rr}$ of
      state $\ket{r}$, the population $\rho_{ee}$ of state $\ket{e}$, the
      probability that there is one photon in the cavity $\rho_{aa}$, and the
      probability that the photon is in the transmission line $P_r$,
      Eq.~(\ref{eq:pr}). The parameters are $(g,\kappa)=(4.9,2.42)\MHz$,
      $\gamma=\kbad=\Delta=0$ and $\Tc=0.009\unit{\mu s}$, thus the regime is
      non adiabatic as $\Tc\approx 0.57/(\gamma C)$. At $t=t_2$ the population
      $\rho_{rr}$ gives $\eta(t_2)$. In this case the system has been simulated
      for a longer time interval: $t_2=-t_1=15\Tc$.} }
\end{figure}
In order to get insight into the optimized dynamics we analyse the time
dependence of the control pulse as well as the dynamics of cavity and atomic
state populations for $\Tc=0.009\unit{\mu s}$, namely, when the dynamics is
non-adiabatic. Figure~\ref{fig:5}(a) shows the time evolution of the pulse
$\Oo(t)$ resulting from the optimization procedure in the non-adiabatic regime;
the pulse $\OX(t)$ is shown for comparison. The \red{efficiency} of the
transfer (when the losses are neglected) with the control pulse $\OX$ is
$\eta^\mathrm{X}\approx 0.07<\eta_\mathrm{max}$ because the process is non
adiabatic, while the \red{efficiency} reached with the optimized pulse $\Oo(t)$
is $\eta^\mathrm{opt}\approx 0.63$. The value of the solid green line at
$t=t_2$ in Fig.~\ref{fig:5}(b) and ~\ref{fig:5}(c) corresponds to the leftmost
point in Fig.~\ref{fig:4} for the case without losses. \red{A double bump in
  the cavity population is visible in Fig.~\ref{fig:5}(b): this is due to the
  Jaynes-Cummings dynamics, and is thus the periodic exchange of population
  between the atomic excited state $\ket{e}$ and the cavity field.} In
Fig.~\ref{fig:5}(a) it is noticeable that the intensity of the optimized pulse
exhibits a relatively high peak when the photon is impinging on the cavity. It
corresponds to a way to perform impedance matching in order to maximize the
transmission at the mirror, \red{and it is related to the same dynamics which
  gives rise to the divergence of $\OF(t)$ and $\OD(t)$ which is found when
  they are applied in the non-adiabatic regime}. After this the intensity of
the control pulse vanishes and then exhibits a second maximum when the
population of the excited state reaches the maximum: we verified that the area
about this second ``pulse'' corresponds to the one of a $\pi$ pulse, thus
transferring the population into state $\ket{r}$.

We now investigate the limit of optimal storage. For this purpose we determine
the lower bound $\Tc^\mathrm{min}$ to the coherence time $\Tc$ of the photon,
for which a given \red{efficiency} $\eta=\eta_\mathrm{tr}$ can be reached. For
each value of $g$ and $\Tc$ we optimize the control pulse using GRAPE. For each
$g$ we determine $\eta$ as a function of $\Tc$ and then extract
$\Tc^\mathrm{min} = \min_{\Tc}\brrr{\Tc: \eta(\Tc)\geq\eta_\mathrm{tr}}$. We
then analyse how the minimum coherence time $\Tc^\mathrm{min}$ scales with the
vacuum Rabi frequency $g$.

\begin{figure}[!ht]
  \centering \includegraphics[]{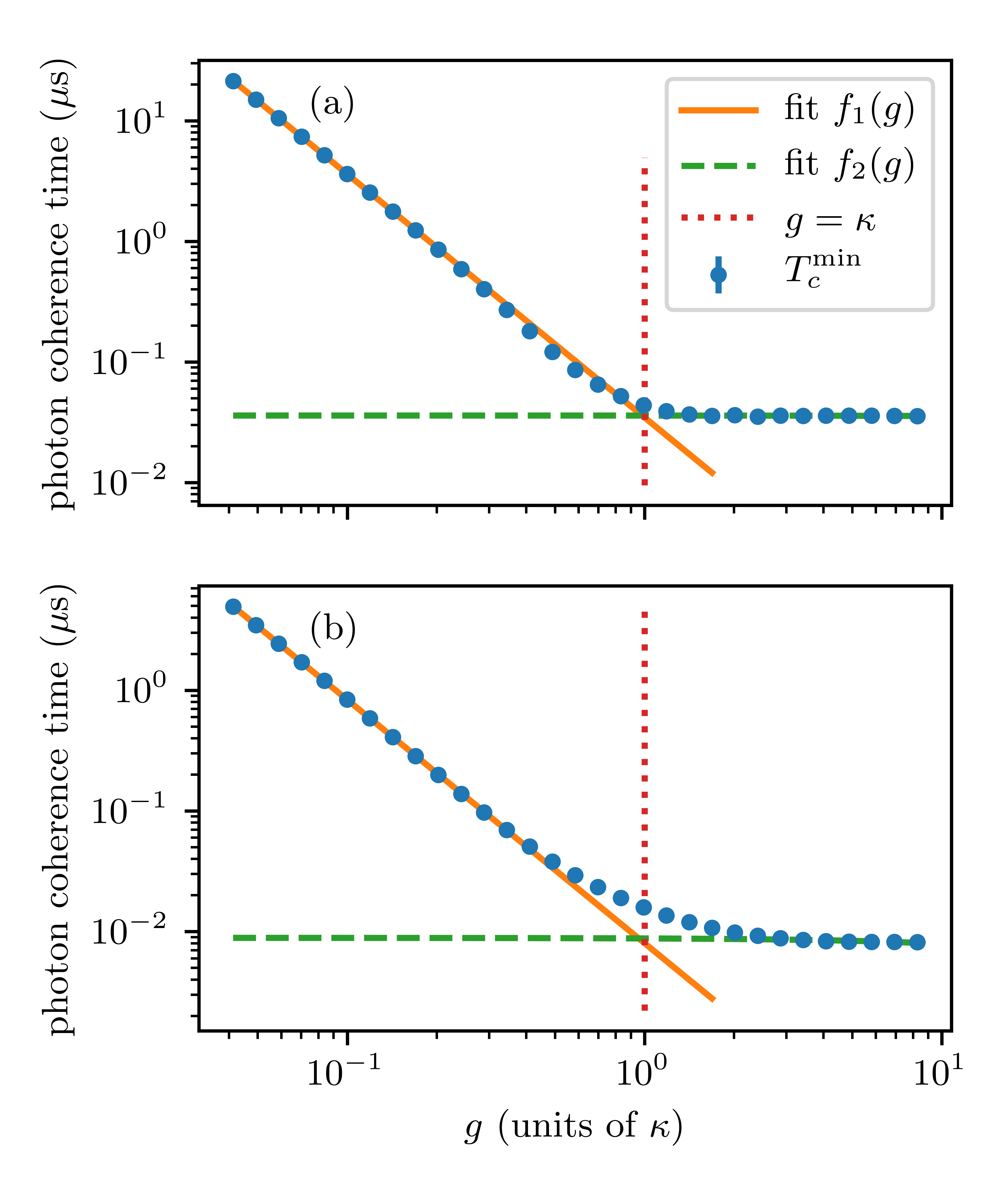}
  \caption{\label{fig:6}Minimum photon coherence time $\Tc^\mathrm{min}$ as a
    function of $g$ (in units of $\kappa$). The coherence time
    $\Tc^\mathrm{min}$ is the lower bound to the coherence time of photons
    which can be stored with \red{efficiency} (a) $\eta_\mathrm{tr}=0.99$ and
    (b) $\eta_\mathrm{tr}=2/3$ for $\gamma=\kbad=\Delta=0$. The vertical dotted
    line shows the value $g=\kappa=2.42\MHz$. The data in the region
    $g\ll\kappa$ and $g\gg\kappa$ have been fitted with the functions $f_1(g) =
    a \kappa/g^2$ and $f_2(g) = a'/ \kappa$, respectively.}
\end{figure}

Figure~\ref{fig:6} displays the minimum photon coherence time
$\Tc^\mathrm{min}$ required for reaching the storage \red{efficiency} (a)
$\eta_\mathrm{tr}=0.99$ and (b) $\eta_\mathrm{tr}=2/3$ as a function of the
coupling constant $g$. We observe two behaviours, separated by the value
$g=\kappa$: For $g\ll\kappa$, in the bad cavity limit, we extract the
functional behaviour $\Tc^\mathrm{min}\propto 1/\gamma C=\kappa/g^2$. On the
contrary, in the good cavity limit, $g>\kappa$, we find that
$\Tc^\mathrm{min}\propto 1/\kappa$: The limit to photon storage is here
determined by the cavity linewidth. The general behaviour as a function of $g$
interpolates between these two limits. This result shows that the photon can be
stored as long as its spectral width is of the order of the linewidth of the
dressed atomic state.

\section{Conclusions}
\label{sec:conclusions}

We have analysed the storage efficiency of a single photon by a single atom
inside a resonator. We have focused on the good cavity limit and shown that, as
in the bad cavity limit, the storage \red{efficiency} is bound by the
cooperativity and the maximal value it can reach is given by
Eq.~\eqref{eq:etamax}. We have extended these predictions to the case in which
the resonator undergoes parasitic losses. For this case we determined the
maximal storage \red{efficiency} for an adiabatic protocol as well as the
corresponding control field respectively given in Eq.~(\ref{eq:etamaxNew}) and
Eq.~(\ref{eq:OX}). Numerical simulations show that protocols based on optimal
control theory do not achieve higher storage \red{efficiencies} than $\eta_{\rm
  max}'$. Nevertheless they can reach this upper bound even for
spectrally-broad photon wave packets where the dynamics is non-adiabatic, as
long as the spectral width is of the order of the linewidth of the dressed
atomic state.

Our analysis shows that the storage efficiency is limited by parasitic losses.
Nevertheless, we have demonstrated that these can be partially compensated by
the choice of an appropriate control field. This result has been analytically
derived for adiabatic protocols, yet it shows that extending optimal control
theory to incoherent dynamics could provide new tools for efficient quantum
memories.

\begin{acknowledgments}
  LG and TS thank the MPI for Quantum Optics in Garching for the kind
  hospitality during completion of this work. The authors acknowledge
  discussions with J\"urgen Eschner, Alexey Kalachev, Anders S. Sørensen,
  Matthias K\"orber, Stefan Langenfeld, Olivier Morin, and Daniel Reich. We are
  especially thankful to Susanne Blum and Gerhard Rempe for helpful comments.
  Financial support by the German Ministry for Education and Research (BMBF)
  under the project Q.com-Q is gratefully acknowledged.
\end{acknowledgments}

\appendix
\section{\label{Appendix:A}Input-output formalism}
In input-output formalism~\cite{Walls1994} the equation of motion are
\begin{equation}
  \label{eq:heiselangev}
  \begin{aligned}
    &\dot{\hat{a}} =-\I g\hat{\sigma}_{ge}- \I\sqrt{2\kappa}
    \hat{\mathcal{E}}_\mathrm{in}(t) - (\kappa+\kappa_\mathrm{bad}) \hat{a}(t)
    + \hat{F}_a, \\
    &\dot{\hat{\sigma}}_{gg} = \I g\hat{\sigma}_{eg}\hat{a}
    -\I g\hat{a}^\dag\hat{\sigma}_{ge}, \\
    &\dot{\hat{\sigma}}_{rr} = \I \Omega(t)\hat{\sigma}_{er}
    -\I \Omega^*(t)\hat{\sigma}_{re}, \\
    &\begin{aligned} \dot{\hat{\sigma}}_{ee} ={}&-\I g\hat{\sigma}_{eg}\hat{a}
      + \I
      g\hat{a}^\dag\hat{\sigma}_{ge} - \I\Omega(t)\hat{\sigma}_{er} + {} \\
      &{}+\I\Omega^*(t)\hat{\sigma}_{re} -\gamma\hat{\sigma}_{ee} +
      \hat{F}_{ee},
    \end{aligned} \\
    &\begin{aligned} \dot{\hat{\sigma}}_{ge} ={} &\I\Delta\hat{\sigma}_{ge} +\I
      g(\hat{\sigma}_{ee}-\hat{\sigma}_{gg})\hat{a}-\I\Omega(t)\hat{\sigma}_{gr}+{} \\
      &{}-\frac{\gamma}{2}\hat{\sigma}_{ge} +\hat{F}_{ge},
    \end{aligned} \\
    &\begin{aligned} \dot{\hat{\sigma}}_{er} = {}&-\I\Delta\hat{\sigma}_{er}
      +\I g\hat{a}^\dag\hat{\sigma}_{gr} +\I\Omega^*(t)(\hat{\sigma}_{rr}
      -\hat{\sigma}_{ee})+ {}\\
      &-\frac{\gamma}{2}\hat{\sigma}_{er}+\hat{F}_{er},
    \end{aligned} \\
    &\dot{\hat{\sigma}}_{gr} = \I g\hat{\sigma}_{er}\hat{a} - \I
    \Omega^*(t)\hat{\sigma}_{ge},
  \end{aligned}
\end{equation}
where $\hat{\sigma}_{jk}=\ketbra{j}{k}$ are atomic operators and $\hat{F}_a$,
$\hat{F}_{ee}$, $\hat{F}_{ge}$ and $\hat{F}_{er}$ are Langevin noise
operators~\cite{Cohen-Tannoudji1994}. The input operator for the quantum
electromagnetic field is
\begin{equation}
  \label{eq:Ein_app}
  \hat{\mathcal{E}}_\mathrm{in}(t) = \sqrt{\frac{Lc}{2\pi^2}}
  \int_{-\infty}^{\infty}e^{-\I kc(t-t_1)}\hat{b}(k+k_c,t=t_1) \de k,
\end{equation}
here $\hat{b}(k,t=t_1)$ is the annihilation operator of the mode $k$ at the
initial time $t=t_1$. The input output relation is given by
\begin{equation}
  \label{eq:inputoutput}
  \hat{\mathcal{E}}_\mathrm{out}(t) = \I\sqrt{2\kappa}\hat{a}(t)
  -\hat{\mathcal{E}}_\mathrm{in}(t).
\end{equation}

The equations of motion for $M\gg1$ atoms in the cavity take the same form as
Eqs.~(\ref{eq:heiselangev}) when one performs the replacement
$\hat{\sigma}_{jk}\to \sum_{i=1}^N\hat{\sigma}_{jk}^i$~\cite{Gorshkov2007a}. In
this case, one can make the approximations $\avg{\tilde{\sigma}_{gg}(t)}\approx
M$, $\avg{\tilde{\sigma}_{rr}(t)}\approx \avg{\tilde{\sigma}_{ee}(t)}\approx
\avg{\tilde{\sigma}_{er}(t)} = 0$, where $\avg{\cdot} = \Tr\br{\rho_0\cdot}$
and $\rho_0$ is the initial state. Then, the set of
equations~(\ref{eq:heiselangev}) reduces to the equations of motion of a single
photon given in Eqs.~(\ref{eq:phi1_inout}).

We note that the quantum impedance matching condition imposed by the authors of
Refs.~\cite{Fleischhauer2000} consists in taking
$\Eout(t)=\dot{\E}_\mathrm{out}(t)=0$, according to which the form of the
control pulse $\OF$, Eq. \eqref{eq:OF}, is found.

\subsection{Effect of photon detuning on storage}
\begin{figure}[!ht]
  \centering \includegraphics[]{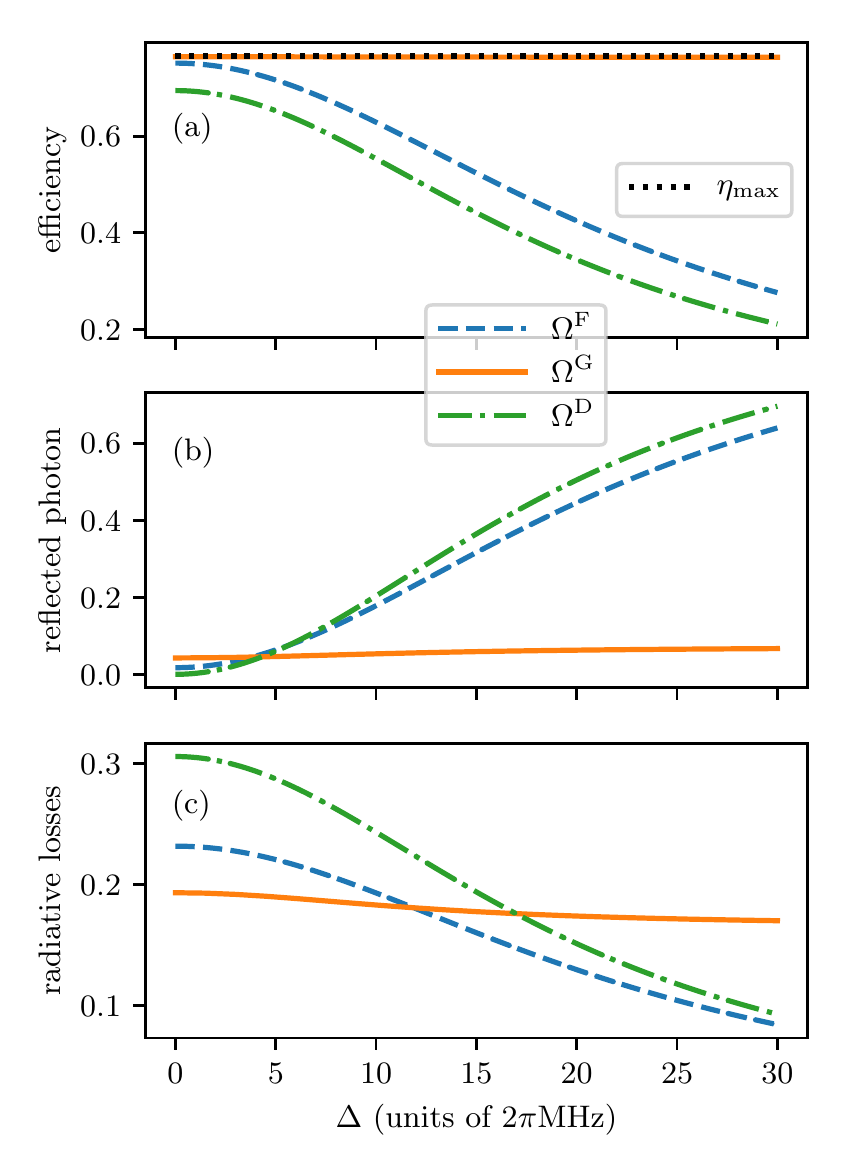}
  \caption{\label{fig:comparisonDelta}(a) Storage \red{efficiency}, (b)
    probability of photon reflection, Eq.~\eqref{eq:pr}, and (c) probability of
    spontaneous decay, Eq.~\eqref{eq:ps}, as a function of the single photon
    detuning $\Delta$ and at time $t_2$. The parameters are
    $(g,\kappa,\gamma)=(4.9,2.42,3.03)\MHz$, $\Tc=0.5\unit{\mu s}$. The input
    photon $\Ein(t)$ is defined as in Eq.~\eqref{eq:Einhypsec}. See Fig.
    \ref{fig:2} for further details.}
\end{figure}
The protocol $\OG(t)$ does not have any restriction on $\Delta$: for every
$\Delta$ there is a pulse $\OG(t)$ that allows for storage with
\red{efficiency} $\eta_\mathrm{max}$ (within the adiabatic regime), see
Eq.~(\ref{eq:OG}). Figure~\ref{fig:comparisonDelta} displays the storage
\red{efficiency} and the losses for each protocol as a function of $\Delta$, as
expected the protocol $\OG(t)$ performs in the same way for any values of
$\Delta$.

A time-dependent phase $\chi(t)$ of the control pulse
$\Omega(t)=\abs{\Omega(t)}e^{\I\chi(t)}$ can be implemented as a two-photon
detuning
\begin{equation}
  \label{eq:twophotdet}
  \delta = \dot{\chi}(t).
\end{equation}
In fact, by applying the unitary transformation
$\hat{U}(t)=\exp\br{-\I\ketbra{r}{r}\chi(t)}$, the transformed Hamiltonian is
$\hat{H}' = \hat{H}'_\mathrm{I} + \hat{H}_\mathrm{fields}$, where
\begin{equation}
  \label{eq:hatprime}
  \begin{aligned}
    \hat{H}'_\mathrm{I}={}&\dot{\chi}(t)\ketbra{r}{r}-\Delta\ketbra{e}{e}+{}\\
    &{}+\br{g\ketbra{e}{g}\hat{a}+\abs{\Omega(t)}\ketbra{e}{r}+\mathrm{H.c.}}.
  \end{aligned}
\end{equation}
For $\OG(t)$ we have
\begin{subequations}
  \begin{align}
    \label{eq:phidotgorsh}
    \dot{\chi}^\mathrm{G}(t)&=\frac{-\Delta}{2\gamma(1+C)}\cdot
                              \frac{\abs{\Ein(t)}^2}
                              {\int_{t_1}^{t}\abs{\Ein(t')}^2\det'}=\\
    \label{eq:dotphiacstarksh}
                            &=\frac{-\Delta\abs{\OG(t)}^2}
                              {\Delta^2+\gamma^2(1+C)^2}\,.
  \end{align}
\end{subequations}
Recall that also $\abs{\OG(t)}$ depends on $\Delta$. This can be understood in
terms of AC Stark shift: one-photon detuning $\Delta\neq0$ is a shift of the
control laser out of resonance for the transition $\ket{r}-\ket{e}$ and thereby
induces an AC Stark shift on the levels $\ket{e}$ and $\ket{r}$ of the atom;
thus the condition of two-photon resonance does not hold anymore. In order to
restore the latter, changes in frequency of the carrier and/or of the cavity
and/or of the atomic levels are needed and they appear as a two-photon detuning
in the Hamiltonian. This also explains why the reflected photon probability for
the protocols $\OF(t)$ and $\OD(t)$ (see Fig.~\ref{fig:comparisonDelta}), which
do not take into account the one-photon detuning, increases with increasing
$\Delta$: the input photon sees the system out of resonance and hence it is
mostly reflected.

Eq.~\eqref{eq:dotphiacstarksh} gives the energy shift as a function of the Rabi
frequency of the control pulse.

\end{document}